\documentclass{aipproc}
\layoutstyle{6x9}
\usepackage{amsmath}
\usepackage[psamsfonts]{amssymb}

\newtheorem{Theorem}{Theorem}

\begin{document}

\title{Geometrical Structures of Space-Time in General Relativity\footnote{
The following article has been accepted by AIP Conference Proceedings. 
After it is published, it will be found at
\emph{http://proceedings.aip.org/proceedings}.
}}


\author{Ignacio S\'anchez-Rodr\'{\i}guez}{
address={Department of Geometry and Topology,
University of Granada,\\
E-18071 Granada, Spain}
}

\begin{abstract}
Space-Time in general relativity is
a dynamical entity because it is
subject to the Einstein field equations.
The space-time metric provides
different geometrical
structures: conformal, volume,
projective and linear connection.
A deep understanding of them
has consequences on the dynamical
role played by geometry.
We present a unified description
of those geometrical structures,
with a standard criterion of
naturalness, and then
we establish relationships among them
and try to clarify the meaning of
associated geometric magnitudes.
\end{abstract}

\maketitle

\section{Introduction}
\hspace{.19 in}The space-time of general
relativity (GR), from the point of view of
differential geometry,
is a $4$-dimensional manifold
$M$, with a $C^{\, \infty }$
atlas $\mathcal{A}$.
The atlas is the
\emph{differential structure}
of our space-time.

The \emph{principle of general covariance} of GR
establishes the invariance
by diffeomorphisms.
This leads us to think that
a \emph{physical event}
is not a point, but
a geometrical
structure on a neighborhood.
The \emph{fundamental
geometrical structures}
that we can consider defined in
the space-time are:
\begin{itemize}
\item Volume ($4$-form)
\item Conformal structure (Lorentzian)
\item Metric (Lorentzian)
\item Linear connection (symmetric)
\item Projective structure
\end{itemize}

They are defined
in terms of the
\emph{most primitive}
differential structure,
via the
concept of $G$-structure.
Volume, conformal structure and metric
are \emph{first
order $G$-structures}.
But linear connection
and projective structure
are \emph{second order
$G$-structures}.

For certain $G$'s,
\emph{classified in}
\cite{KoNa},
every first order $G$-structure
lead to a unique second order
structure, named its
\emph{prolongation}.
This is the case for
the volume, metric and conformal
structures.

\section{Frame bundles}
\hspace{.19 in}The \emph{$r$-th order
frame bundle} $\mathcal{F}^rM$ is
a quotient space of a subset of
$\mathcal{A}$
\cite[p. 38]{Tesis}.
An \emph{$r$-frame}, $j^r\varphi
\in \mathcal{F}^rM$
is an $r$-jet at
$0$, where $x=\varphi ^{-1}$
is a chart
with $0$ as a target.

We restrict our interest
to first and second order.
The first order frame
bundle $\mathcal{F}^1M$ is
usually identified with the
\emph{linear frame bundle}
$L\, M$.
For a better
understanding of the second
order frame bundle
consider $L\, L\, M$, the linear
bundle of $L\, M$. There is a
\emph{canonical inclusion}
$\mathcal{F}^2M
\hookrightarrow L\, L\, M$,
$j^2\varphi \mapsto
j^1\tilde{\varphi }$,
where $\tilde{\varphi }$
is the diffeomorphism
induced by $\varphi $,
between neighborhoods
of $0\in \mathbf{R}^{n+n^2}$
and $j^1\varphi \in L\, M$
\cite[p. 139]{KobTG}.

Let $J^1L\, M$ be the bundle
of \emph{$1$-jets
of (local) sections}
of $L\, M$ and $s$ be a section of $L\, M$.
Each $j^1_p\, s$
is characterized by the
\emph{transversal
$n$-subspace}
$H_l= s_{\ast}(T_p\, M)
\subset T_{l}\, L\, M$ \cite{PLGar}.
Then, there is also a
\emph{canonical inclusion}
$J^1L\, M\hookrightarrow L\, L\, M$,
$j^1_p\, s\mapsto z$,
where $z$ is the basis
of $T_{l}\, L\, M$, whose first
$n$ vectors span $H_l$
and correspond to the
usual basis
of $\mathbf{R}^n$,
via the \emph{canonical form
of $L\, M$}, and the last
$n^2$ vectors are the
fundamental vectors
corresponding
to the standard basis of
$\mathfrak{gl}(n\, ,\mathbf{R})$
\cite{KoNo}.

By the previous canonical maps, it happens that
$\mathcal{F}^2M$
is mapped one to one
into the subset of $J^1L\, M$,
corresponding with the \emph{torsion-free
transversal $n$-subspaces} in $TL\, M$.
\begin{Theorem}
We have the canonical embeddings: \emph{\cite[p. 54]{Tesis}}
$$ \mathcal{F}^2M\ \hookrightarrow \
J^1L\, M\ \hookrightarrow \ L\, L\, M$$
\end{Theorem}

The $r$-th order
frame bundles
are principal bundles. They
are fundamental because
every \emph{natural bundle},
in the categorial approach,
can be described as an associated
bundle to some $\mathcal{F}^rM$
\cite{PaTe} and the so-called
\emph{geometrical
objects} can be identified with
sections of those associated bundles.

\section{Structural groups}
\hspace{.19 in}The structural group
of the principal bundle
$\mathcal{F}^rM$ is
the group
$\mathrm{G}^{\, r}_n$
of \emph{$r$-jets
at 0 of diffeomorphisms
of $\mathbf{R}^n$},
$j^r_0\, \phi $,
with $\phi(0)=0$.

The group
$\mathrm{G}^1_n$ is
identified
with $\mathrm{GL}(n\, ,\mathbf{R})$.
Then there is a canonical
inclusion of
$\mathrm{G}^1_n$
into $\mathrm{G}^{\, r}_n$,
if we take the $r$-jet
at 0 of every
linear map of $\mathbf{R}^n$.
Furthermore, $\mathrm{G}^{\, r}_n$ is the
\emph{semidirect product}
of $\mathrm{G}^1_n$ with a nilpotent
normal subgroup \cite{Terng}.
Let us see this decomposition for
 $\mathrm{G}^2_n$.
We consider the underlying additive group
of the vector space $\mathrm{S}^2_n$
of symmetric bilinear maps
of $\mathbf{R}^n\times \mathbf{R}^n$
into $\mathbf{R}^n$. Then
there is a monomorphism
$\imath \colon \mathrm{S}^2_n
\rightarrow
\mathrm{G}^2_n$ defined
by $\imath (s)=j^2_0\, \phi $,
with $s=(s^{i}_{jk}\, )$ and
$\phi (u^{i}\, ):=
(u^{i}+\frac{1}{2}s^{i}_{jk}u^{j}
u^{k}\, )$.

\begin{Theorem}
We obtain the split
exact sequence of groups:
$$ 0\ \rightarrow \ \mathrm{S}^2_n
\ \stackrel{\imath }{\rightarrow }
\ \mathrm{G}^2_n\
\ \underset{\scriptscriptstyle
\smash[b]{\mathbf{\supset}} }
{\smash[b]{\rightleftarrows }}\
\ \mathrm{G}^1_n\rightarrow \ 1
$$
\end{Theorem}

It makes
\emph{$\mathrm{G}^2_n$
isomorphic to
the semidirect product}
$\mathrm{G}^1_n
\rtimes \mathrm{S}^2_n$,
whose multiplication rule is
$(a\, ,s)(b\, ,\, t):=(ab\, ,\, b^{-1}s(b,b)+t)$.
The isomorphism is given by
$j^2_0\, \phi \mapsto
(D\phi |_{0}\,,\,
D\phi |^{\scriptscriptstyle -1}_0
\, D^2\phi |_0)$.

\section{$G$-structures}
\hspace{.19 in}We define an $r$-\emph{th order
 $G$-structure} on $M$ as
a \emph{reduction}
of $\mathcal{F}^rM$ to a subgroup
$G\subset \mathrm{G}^{\, r}_n$
\cite{KobTG}.
This idea of
\emph{geometrical structure} on
$M$ concerns the classification
of charts in $\mathcal{A}$,
\emph{when the meaningful
classes are chosen} guided by an
structural group.

We exemplify the concept of
a $G$-structure studying a
volume on a manifold,
which rarely is treated
this way \cite{CrSa}.
Let us define a \emph{volume} on $M$ as a
first order $G$-structure $V$,
with $G= \mathrm{SL}^{
\scriptscriptstyle \smash[t]{\pm}}_n:=\{a
\in \mathrm{GL}(n\, ,\mathbf{R}) : \
\mid \det a\mid \, =1\}$.
For an orientable $M$,
$V$ has two components for two $\mathrm{SL}(n
\, , \mathbf{R})$-structures,
for two equal,
except sign, \emph{volume $n$-forms}.
For a general $M$, volume
corresponds to
\emph{odd type $n$-form},
as in \cite[pp. 21-27]{DeRham}.

From \emph{principal bundle
theory} \cite{KoNo},
$\mathrm{SL}^{\scriptscriptstyle \pm}_n$-structures
are the sections of the
\emph{bundle associated} with
$L\, M$ and the left action of
$\mathrm{G}^1_n$ on
$\mathrm{G}^1_n/
\mathrm{SL}^{\scriptscriptstyle \pm}_n$.
This is the
\emph{volume bundle},
$\mathcal{V}M$.
Furthermore, the
sections of $\mathcal{V}M$
correspond to
$\mathrm{G}^1_n$-\emph{equivariant
functions} $f$
of $L\, M$ to $\mathrm{G}^1_n/
\mathrm{SL}^{\scriptscriptstyle \pm}_n$.
The equivariance
condition is
$f(la)=\mid \det a\mid ^{-1/n}I_n\cdot f(l)$,
$\forall a\in
\mathrm{G}^1_n$.

We have the bijections:
$$\text{Volumes on }M
\ \longleftrightarrow \
\operatorname{Sec}
\mathcal{V}M
\ \longleftrightarrow \
C^{\, \infty}_{\text{eq}}
(L\, M\, ,\, \mathrm{G}^1_n/
\mathrm{SL}
^{\scriptscriptstyle \pm}_n)
$$

The isomorphisms $\mathrm{G}^1_n/
\mathrm{SL}^{\scriptscriptstyle \pm}_n
\, \simeq \, \mathrm{H}_n$,
with $\mathrm{H}_n:=
\{ k\, I_n\ :\ k>0\} $
and $\mathrm{H}_n
\, \simeq \, \mathbf{R}^{+}$,
the multiplicative group of positive numbers,
allow to represent a volume
as an \emph{(odd) scalar density}
on $M$.

\section{Second order structures}
\hspace{.19 in}We can view a \emph{symmetric
linear connection} (SLC) on $M$ as a
$\mathrm{G}^1_n$-structure of
second order.
A SLC is also the image of an
\emph{injective homomorphism}
of $L\, M$ to $\mathcal{F}^2M$ \cite{KobTG}.

From the \emph{principal bundle
theory} \cite{KoNo}, SLC's on $M$
are sections of the
\emph{SLC bundle},
$\mathcal{D}M$, associated
with $\mathcal{F}^2M$
and the action of
$\mathrm{G}^2_n$ on
$\mathrm{G}^2_n/
\mathrm{G}^1_n
\, \simeq \, \mathrm{S}^2_n$.
Furthermore, each
SLC, $\nabla $,
corresponds to a
$\mathrm{G}^2_n$-\emph{equivariant
function}
$f^{\, \scriptscriptstyle \nabla }\colon
\mathcal{F}^2M
\rightarrow
\mathrm{S}^2_n$, verifying
$f^{\, \scriptscriptstyle \nabla }
(z(a,s))=a^{-1}f^{\,
\scriptscriptstyle
\nabla }(z)(a,a)+s$.

We have the bijections:
$$\text{SLC's on }M
\quad \longleftrightarrow \quad
\operatorname{Sec}
\mathcal{D}M
\quad \longleftrightarrow \quad
C^{\, \infty}_{\text{eq}}
(\mathcal{F}^2M\, ,\, \mathrm{S}^2_n)
$$

Given two SLC's, $\nabla $
and $\widehat{\nabla }$,
\emph{the difference function} $f^
{\scriptscriptstyle \nabla}
-f^{\, \scriptscriptstyle
\widehat{\nabla }}\colon
\mathcal{F}^2M
\rightarrow
\mathrm{S}^2_n$
verifies $z(a,s)\mapsto
a^{-1}(f^{\, \scriptscriptstyle
\nabla}(z)
-f^{\, \scriptscriptstyle
\widehat{\nabla }}(z))(a,a)$.
Then, it is projectable
to a function
$f\colon L\, M\rightarrow
\mathrm{S}^2_n$
verifying
$f(la)=a^{-1}f(l)(a,a)$,
which corresponds to \emph{a
tensor}
$\rho =(\, \rho ^i_{jk}\, )$ on $M$.

A \emph{projective structure}
(PS) is an equivalence class of SLC's
which have the same family
of \emph{pregeodesics}.
This is the cornerstone
to understand the
\emph{freely falling bodies}
in GR \cite{EPS}.
We can define a PS on $M$
as a \emph{second order
$\mathrm{G}^1_n\rtimes
\mathfrak{p}$-structure},
$Q$, with
$\mathfrak{p}:=\{
s\in \mathrm{S}^2_n\ :\
s^{i}_{jk}=\delta^i_j\mu _k+\mu _j
\delta^i_k,\ \mu =(\mu _i)\in
\mathbf{R}^{n\,\ast}\} $.

Now, for two SLC
\emph{included} in
the same PS (i.e. literally
$\nabla ,\widehat{\nabla }\subset Q$)
the tensor $\rho $, expressing
\emph{their difference},
is determined by
the contraction
$C(\rho )=(\, \rho ^s_{si}\, )$,
which is \emph{an 1-form} on $M$.

\section{Prolongations}
\hspace{.19 in}Let $B$ be a \emph{first
order $G$-structure}.
A connection in $B$
is a distribution $H$ of
transversal $n$-subspaces,
$H_l\subset T_lB$.
If the subspaces are
\emph{free-torsion},
these determine a
\emph{second order
$G$-structure}, whose
$G^1_n$-\emph{extension}
\cite[p. 206]{GHV}
is a SLC on $M$.
Then, we say that $B$
\emph{admits a SLC}.
Let us give two examples:
\begin{itemize}
	\item A SLC
and a parallel volume is an
\emph{equiaffine
structure} on $M$
\cite{NoSa}; hence,
it is a
second order
$\mathrm{SL}^{
\scriptscriptstyle
\pm}_n$-structure.
\item A SLC compatible
with a conformal structure
is a \emph{Weyl
structure}; hence,
it is a
second order
$\mathrm{CO}(n)$-structure
\cite{CoKo}.
\end{itemize}

For a linear group $G$,
let $\mathfrak{g}$ denote
the Lie algebra of $G$.
The \emph{first prolongation
of} $\mathfrak{g}$ is
defined by
$\mathfrak{g}_1:=
\mathrm{S}^2_n\, \cap \, L(
\mathbf{R}^n,\mathfrak{g})$.
We obtain that
$G\rtimes
\mathfrak{g}_1$
\emph{is a subgroup} of
$\mathrm{G}^1_n
\rtimes \mathrm{S}^2_n$,
and hence, a subgroup
of $\mathrm{G}^2_n$
(see more details in \cite{Braga}).

\begin{Theorem}
Let $B\subset L\, M$ be a
$G$-structure, admitting
a SLC. Then, the set
of 2-frames, corresponding
with torsion-free
transversal $n$-subspaces
which are included in $TB$,
is a \emph{reduction of
$\mathcal{F}^2M$ to}
$G\rtimes
\mathfrak{g}_1$. It is named the
\emph{prolongation
of} $B$ and denoted by $B^2$
\emph{(for a proof, see
\cite[pp. 150-155]{Tesis})}.
\end{Theorem}

Let us give a well known example:
if $B$ is an $O(n)$-structure,
$B^2$ is isomorphic to
$B$ on account of
$\mathfrak{o}(n)_1=\{ 0\}$;
this explain the uniqueness
of Levi-Civita connection.

There is an important theorem
\cite{KoNa}
\emph{classifying
the groups} $G$ such
that \emph{every $G$-structure
admits a SLC}: only
the groups of
\emph{volume,
metric and conformal}
structures, and a class of
groups preserving an
1-dimensional distribution,
have this property.

\section{Concluding remarks}
\hspace{.19 in}We have done a unified description
of the geometrical
structures that have
been used by GR to define intrinsic
properties of the space-time.
The unifying criterion, we used
for it,
\emph{not only is natural
in the sense
that geometric objects
are sections of bundles
associated with
the $\mathcal{F}^rM$
frame bundles} \cite{Terng},
\emph{but also in the sense
that the objects themselves
are reductions of
$\mathcal{F}^rM$}.
Therefore,
we have not considered
a \emph{linear connection
with torsion}
because it is a
section of an
associated bundle
of $\mathcal{F}^2M$,
but not a reduction.

We have tried to clarify
the relationships between
the structures involved.
Only simple relations,
such as intersection, inclusion,
reduction and extension,
have been used for it,
on account of the previous
\emph{prolongation}
of $G$-structures admitting SLC.
For instance, it follows readily
from the last section that
the classical \emph{equiaffine} or
\emph{Weyl structures}
can be defined
as the intersection of a SLC
with the prolongation
of a volume or a conformal
structure, respectively.

Recently, some of my research \cite{Opava}
have been taken into consideration
for one of the lines of thought about
quantum gravity \cite{Stachel}.
This contribution is a set of
my latest reflections and conclusions
about geometrical structures
with an eye on the applications
to physics.

\begin{theacknowledgments}
\hspace{.19 in}The author would like to thank
the IFWGP'07 organizers
for their kind invitation to
attend this Workshop
and to present this communication.
I wishes to acknowledge gratefully
Dr. John Stachel the attention
paid to my work.

This work has been partially
supported by the Junta de
Andaluc\'{\i}a P.A.I.: FQM-324.
\end{theacknowledgments}

\bibliographystyle{aipproc}

\end{document}